# Molecular Phylogeny of Chinese Thuidiaceae with emphasis on *Thuidium* and *Pelekium*

QI-YING, CAI[1,2], BI-CAI, GUAN[2], GANG, GE[2], YAN-MING, FANG[1]

[1] College of Biology and the Environment, Nanjing Forestry University, Nanjing 210037, China.

[2] College of Life Science, Nanchang University, 330031 Nanchang, China. E-mail: caiqiying@ncu.edu.cn

**Abstract**

We present molecular phylogenetic investigation of Thuidiaceae, especially on *Thudium* and *Pelekium*. Three chloroplast sequences (trnL-F, rps4, and atpB-rbcL) and one nuclear sequence (ITS) were analyzed. Data partitions were analyzed separately and in combination by employing MP (maximum parsimony) and Bayesian methods. The influence of data conflict in combined analyses was further explored by two methods: the incongruence length difference (ILD) test and the partition addition bootstrap alteration approach (PABA). Based on the results, ITS 1& 2 had crucial effect in phylogenetic reconstruction in this study, and more chloroplast sequences should be combinated into the analyses since their stability for reconstructing within genus of pleurocarpous mosses. We supported that Helodiaceae including *Actinothuidium*, *Bryochenea*, and *Helodium* still attributed to Thuidiaceae, and the monophyletic Thuidiaceae *s. lat.* should also include several genera (or species) from Leskeaceae such as *Haplocladium* and *Leskea*. In the Thuidiaceae, *Thuidium* and *Pelekium* were resolved as two monophyletic groups separately. The results from molecular phylogeny were supported by the crucial morphological characters in Thuidiaceae *s. lat.*, *Thuidium* and *Pelekium*.

## Introduction

Pleurocarpous mosses consist of around 5000 species that are defined by the presence of lateral perichaetia along the gametophyte stems. Monophyletic pleurocarpous mosses were resolved as three orders: Ptychomniales, Hypnales, and Hookeriales (Shaw *et al*. 2003). Because of the rapid diversification that occurred during early stages of pleurocarpous mosses evolution (Shaw *et al*. 2003; Newton *et al*. 2007), the phylogenetic relationships within them, especially in Hypnales, were inadequately investigated through phenotypic characteristics and sequence information.

The traditional classifications are complicated by high levels of homoplay in many morphological characters (Hedenäs 1995;1998) which suffered from numerous reversals or reductions that tend to obscure the evolutionary significance of the original autapomorphy (Hedenäs 1999; Buck *et al*. 2000). Hedenäs (1994, 1995) explored morphological diversity for reconstructing the phylogeny of pleurocarpous mosses. Though some shortcomings presented in these researches, the general conclusions were supported by molecular phylogeny (Buck *et al*. 2000; Troitsky *et al*. 2007). Alternatively, for the rapid diversification, the molecular evolution in pleurocarpous mosses was much slower than acrocarpous mosses. It seemed difficult to resolve the problems through molecular approach. It was suggested that the rapid diversification and evolution may not entirely account for the lack of resolution among families of the Hypnidae (Buck *et al*. 2000). Troitsky *et al*. (2007) also indicated that Shaw's conclusion of rapid diversification resulted only from analyzing distinct regions of the chloroplast genome (sometimes in combination with mitochondrial genes). By combining nuclear ribosomal internal transcribed spacers (ITS) into analyses, the resolutions were improved significantly (Vanderpoorten *et al*. 2002; Gardiner *et al*. 2005; Troitsky *et al*. 2007; Olsson *et al*. 2009; Huttunen et al. 2012; Huttunen et al. 2013). But so far, the contribution of chloroplast genome and ITS 1& 2 for reconstructing relationship of pleurocarpous mosses has not been assessed quantitatively. Matthew et al. (2016) presented the first phylogenetic inference from high-throughput sequence data (transcriptome sequences) for pleurocarpous mosses. A large amount of new information can be extracted through this approach, but relatively high costs temporarily limit its promotion.

    The Thuidiaceae is a small, cosmopolitan family with its centre of diversity in Asia. Opinions vary regarding familial limits. Thuidiaceae and Leskeaceae are two closely related families in Hypnales. Schimper (1876) established the tribus Leskeaceae, under which *subfam*. Thuidieae was included. Kindberg (1897) first established Thuidiaceae including *Myurella*, *Heterocladium*, *Pseudoleskeella*, *Pseudoleskea*, and *Thuidium*. Since then, the Thuidiaceae has been placed close to Leskeaceae in most traditional taxonomies (i.e. Kindberg 1897; Fleischer 1904–1923; Brotherus 1925; Crum & Anderson 1981; Buck & Crum 1990). Similar characters present repeatedly in different genera and species of the two families, and the circumscription through phenotype was subjective. So the discussions about the taxonomy of Thuidiaceae and Leskeaceae go on. Latter results from molecular systematics studies (Gardiner *et al*. 2005; Troitsky *et al*. 2007; Garcíń-Avila *et al*. 2009) also demonstrated the close relationship that is staggered in the phylogenetic trees. As an example, *Leskea polycarpa* shows the characters of the Leskaceae, however was located within the Thuidiaceae (Troitsky *et al*. 2007; Garcíń-Avila *et al*. 2009). The relationships inbetween and within Thuidiaceae and Leskeaceae seem to be more complicated, but more distinct. So far, the monophyly of the two families defined by traditional classifications, and the latter three-mosses classification (Buck & Goffinet 2000, Goffinet & Buck 2004, Goffinet *et al*. 2008), has been challenged by the recent evidences.

    Much attention has been focused on the taxonomy of Thuidiaceae in broad sense, since Brotherus (1925) following Fleischer (1904–1923) redefined Thuidiaceae as including four subfamilies, i.e. Anomodontoideae, Helodioideae, Heterocldioideae, and Thuidioideae. Some taxonomies were showed in table 1, in which adjustments were made on the level of genus, subfamily, and even family. Watanabe (1972) combined Helodioideae into Thuidioideae, and moved *Haplocladium* and *Claopodium* from Anomodontoideae into Thuidioideae. *Haplocladium*, *Anomodon*, and *Herpetineuron* were removed from Thuidiaceae and attributed to Leskeaceae (Crum & Anderson 1981). Buck & Crum (1990) further established Helodiaceae contains only one genus *Helodium*; and moved *Bryonoguchia* to Thuidioideae, and *Hylocomiposis* and *Actinothuidium* attributed to Hylocomiaceae respectively. They also described a new subfamily Cyrtohypnoideae to house some groups previously related to *Thuidium s. str.*; and redefined Anomodontoideae as a family. As for Chinese Thuidiaceae, results from Wu & Jia (2000) and Wu *et al*. (2002) had been consistent with those from Buck & Crum (1990) in the classification of Anomodontaceae, however they included Heterocladioideae, Thuidioideae, and Helodioideae into Thuidiaceae and *Abietinella* to Helodioideae. *Haplocladium* and *Claopodium* were still retained in

Thuidioideae. Similar classification pattern was also adopted by Fang & Koponen (2001). In other researches (Buck & Goffinet 2000, Goffinet & Buck 2004, Goffinet et al. 2008), Helodiaceae was still established, which includes *Actinothuidium*, *Bryochenea*, and *Helodium*.

The phylogeny of Thuidiaceae is not fully resolved with substantial support. The ambiguous and conflicting results may be the result of the use of few species, too little sequences data, phylogenetic methods that do not adequately capture the complex nature of DNA evolution. We selected four most frequently used regions (Stech & Quandt, 2010): plastid *trnL-F*, *atpB-rbcL*, *rps4*, and nuclear ITS 1& 2 for phylogenetic reconstruction in attempt to: 1) circumscribe a monophyletic Thuidiaceae in pleurocarpous mosses; 2) reconstruct well supported phylogenetic relationships inbetween and within *Thuidium* and *Pelekium*; 3) detect conflict or incongruence between different data sets for accessing reliability of the reconstruction, from which the more reasonable strategies for subsequent reconstruction will be provided.

MATERIAL AND METHODS

*Taxa sampling*

54 species from the Order Hypnales, most of them belong to Thuidiaceae and Leskeaceae (Goffinet et al. 2008) were collected from ChangBai Mountain (JiLin & LiaoLing), LongQuan Mountain (ZheJiang), Huangshan Mountin (AnHui), Lushan Mountain (LS, JiangXi), ShenNongJia (SNJ, HuBei), Jade Dragon Snow Mountain (LJ, YunNan) and XiShuangBanna (YunNan) of China. Voucher specimens were identified. Another 19 species were also included into the analyses, and their accession numbers in Genebank were listed in table 2. Of all the sampled species, seventy- two were ingroup species belonging to Hypnales and one outgroup species, *Hookeria acutifolia*, belongs to Hookeriales.

*Nucleotide sampling and laboratory procedures*

The dried specimens were saturated with DDW and cleaned under stereo microscope. The leafy gametophytes were blotted up with qualitative filter paper and stored at -70°C. Genomic DNA was extracted using Universal Genomic DNA Extraction Kit Ver. 3.0 (TaKaRa Code: DV811A). Four fragments (cpDNA: *trnL-F*, *rps4*, and *atpB-rbcL*, and nuclear ITS) were amplified and sequenced. The PCR protocols were listed blow.

*trnL-F*. Primer pairs trnL–F (C) (5′-CGA AAT CGG TAG ACG CTA CG-3′) / trnL-F (F) (5'-ATT TGA ACT GGT GAC ACG AG-3') (Taberlet et al. 1991). Reaction mixture 50μl: 10×Ex Taq Buffer ($Mg^{2+}$ Free) 5.0μl, dNTP Mixture (each 2.5 mM) 4.0μl, $MgCl_2$ (25 mM) 4.0μl, each primers (10μM) 2.0μl, DMSO 2.5μl, non-diluted genomic DNA (~50ng/μl) 1.0μl, *TaKaRa Ex Taq*® (5 U/μl) (Takara Bio. Inc.) 0.25μl. Procedure: 1 cycle (2 min 94 °C), 35 cycles (1 min 94 °C, 1min 52 °C, 1 min 72 °C), 1 cycle (7 min 72 °C).

*rps4*. Primer pairs *trnS*-F (5′-TAC CGA GGG TTC GAA TC-3′) (Souza-Chies et al. 1997) / *rps* 5 (5′-ATG TCC CGT TAT CGA GGA CCT-3′) (Nadot et al. 1994). The reaction mixture and procedure were same as trnL−F.

*atpB-rbcL*. Primer pairs *atpB*-1 (5′-ACATCKARTACKGGACCAATAA-3′) / *rbcL*-1(5′-AACACCAGCTTTRAATCCAA-3′) (Chiang et al. 1998). Reaction mixture 50μl: 10×*Ex Taq* Buffer ($Mg^{2+}$ Free) 5.0μl, dNTP Mixture (each 2.5 mM) 4.0μl, $MgCl_2$ (25 mM) 4.0μl, each primers (10μM) 2.0μl, DMSO 4.0μl, non-diluted genomic DNA (~50ng/μl) 1.0μl, *TaKaRa Ex Taq*® (5 U/μl) (Takara Bio. Inc.) 0.25μl. Procedure: 1 cycle (2 min 94 °C), 35 cycles (1 min 94 °C, 75sec 54 °C, 75sec 72 °C), 1 cycle (7 min 72 °C).

ITS. Primer pairs ITS4-bryo (5′-TCC TCC GCT TAG TGA TAT GC-3′)/ ITS5-bryo (5′-GGA AGG AGA AGT CGT AAC AAG G-3′) (Stech & Frahm 1999). Reaction mixture 50μl: 10×*Ex Taq* Buffer ($Mg^{2+}$ Free) 5.0μl, dNTP Mixture (each 2.5 mM) 4.0μl, $MgCl_2$ (25 mM) 3.75μl, each primers (10μM) 2.0μl, DMSO 4.0μl, non-diluted genomic DNA (~50ng/μl) 1.0μl, *TaKaRa Ex Taq*® (5 U/μl) (Takara Bio. Inc.) 0.25μl. Procedure: 1 cycle (2 min 94 ℃), 35 cycles (1 min 94 ℃, 75sec 56 ℃, 75sec 72 ℃), 1 cycle (7 min 72 ℃).

All products were verified on a 1% agarose gel and purified with Agarose Gel DNA Purification Kit (Takara) and sequenced directly with the same primers used for PCR reaction by Invitrogen™ (ShangHai).

*Sequence alignment and phylogenetic analyses*

Clustal X 2.0.11 (Thompson *et al.* 1997) was used to performed sequence alignment with default parameters. BioEditor 7.0.9 (Hall 2005) was used to exclude ambiguous alignment positions. FastGap 1.2 (Borchsenius 2009) was used to code gap or indel characters as binary characters (A or C) using the simple method of Simmons & Ochoterena (2000) and add them to the data file as separate partitions. The results from FastGap are no different from those obtained with GapCoder (Young & Healy 2003).

Maximum parsimony (MP) analyses were performed using PAUP 4.0 b 10 (Swofford 2002) for both separated and combined data sets. Heuristic searches were performed with 1000 random taxa addition replicates and TBR branch swapping. Bootstrap values analyses were performed for each sequence separately and for each combined date set, as well. All trees were displayed with TREEVIEW 1.6.6 (Page 1996). Bayesian inference was employed with MrBayes 3.2 (Huelsenbeck & Ronquist 2001; Ronquist & Huelsenbeck 2003). Bayesian inference criterion (BIC) of each DNA fragment was calculated respectively within jModltest 0.1.1 (Guindon S & Gascuel O. 2003; Posada D. 2008). Then partitioned analyses of different data sets with the models of their own were set up following MrBayes 3.2 Manual (Ronquist *et al.* 2007). Metropolis-coupled Markov chain Monte Carlo sampling was performed with four chains. It ran 5 million generations for combination of three cpDNA, 7 million generations for combination of all sequences, and 10 million generations for the combination of all date sets before the standard deviation of split frequencies was below 0.01.

*Data incongruence analyses*

ILD test (Farris *et al.* 1995) and PABA approach (Struck *et al.* 2006) were conducted to detect phylogeny data incongruence. The ILD test, also known as the "partition homogeneity test" in PAUP, was employed to measure character conflicts under a parsimony framework between each two data sets using 1000 heuristic search repetitions, TBR branch swapping, and simple taxon addition for all combined data sets.

A partition addition bootstrap alteration (PABA) approach (Struck *et al.* 2006, Struck 2007) was employed to indentify the node and/or data partition causing the incongruence by examining methodologically the alteration (δ) of bootstrap support at each given node when additional data partitions are added. Thirty nodes (shown with black spots) in MP tree (Fig. 3) reconstructed on all four date sets were selected to be further analyzed (shown in Table 4 and 5).

RESULT

*Data characteristics*

Of the 4765 characters (trnL−F: 579, rps4: 657, atpB−rbcL: 1011, ITS: 2518), 1257 (trnL−F:

100, rps4: 107, atpB−rbcL: 226, and ITS: 824) sites were parsimony informative. The BICs for sequences were HKY+G (trnL−F), TPM3uf+G (rps4), TPM1uf+G (atpB-rbcL), and GTR+G (ITS) respectively.

*Phylogenetic analyses*

*Combination of cpDNA*

Based on the analyses of three chloroplast sequences (trnL−F, rps4, and atpB−rbcL) a Bayesian inference tree (BI tree) (Fig. 1) and a maximum parsimony tree (MP tree) (2011 steps, CI: 0.5554, HI: 0.4446) were produced. Within the tree a monophyletic clade (i.e. node 16 in Fig. 3) including *Abietinella*, *Actinothuidium*, *Bryonoguchia*, *Cyrto-hypnum*, *Haplocladium*, *Helodium*, *Leskea*, *Platylomella*, *Rauiella*, and *Thuidium* was indicated (bootstrap support (bs):77, Bayesian posterior probabilities (bpp): 1.00). Most taxa from this group had been attributed or related to Thuidiaceae in previous research (Brotherus 1925; Watanabe 1972; Wu 2002). Therefore this group was defined as Thuidiaceae *s. lat*. *Thuidium* and *Pelekium* were separately resolved as monophyletic group, and the basic relationships among species in these two groups were revealed also. *Rauiella fujisama*, *Bryonoguchia molkenboeri* and *Abietinella abietina* formed a clade (i.e. node 13, bs: 18, bpp: 0.50). *Haplocladium* species together with *Platylomella lescurii* and *Helodium paludosum* formed another clade (i.e. node 12, bs: 5, bpp: 0.95) as the basal group of Thuidiaceae *s. lat*. Leskeaceae genera except *Leskea* and *Haplocladium* were resolved as a nonmonophyletic group including *Claopodium*, *Leptopterigynandrum*, *Lescuraea*, *Leskeella*, *Lindbergia*, and *Rozea* in this study.

*Combination of cpDNA and nuclear ITS*

Analyses of three chloroplast sequences (trnL-F, rps4, atpB-rbcL) and nuclear ITS 1& 2 resulted in a maximum parsimony tree (MP tree) (Fig. 2, Tree length: 5736 CI: 0.5281, HI: 0.4719) and a Bayesian inference tree (BI tree) (Fig. 3). Compared with figure 1, the resolution of the phylogenetic relationships was improved evidently, and the backbone of Hypnales was established with high resolution. Two trees resulted from this study exhibited the same structure in the arrangement of 27 selected nodes.

Four main clades in Thuidiaceae *s*. *lat*. were established respectively at node 4, 8, 12, and 14. In the *Thuidium* clade (i.e. node 4), *Thuidium pristocalyx* was resolved as the basal group, and two stable groups (*T. assimile–T. cymbifolium* and *T. kanedae–T. submicropteris*) emerged in all topologies, and two *T. phillbertii* were always located at the top of *Thuidium* clade. However, the relationships within *Thuidium* seemed inconsistent, especially the phylogenetic position of *T. subglaucinum.* The relationships inbetween and within *Pelekium* clade (i.e. nod 8) were consistent relatively. In this clade, *Pelekium tamariscellum*, *P. sparsifolium*, and *P. contortulum* united the basal clade (at node 7, bs: 100, bpp: 1.00), and other species united another clade (at node 6, bs: 92, bpp: 1.00). The third clade (at node 12) also presented like in combination of cpDNA, and *Helodium paludosum* and *Haplocladium virginianum* were united as basal group of this clade. *Actinothuidium hookeri*, *Rauiella fujisana*, *Bryonoguchia molkenboeri*, *Abietinella abietina*, and *Helodium paludosum* united a monophyletic clade (at node 14, bs: 41, bpp: 0.98). *Claopodium*, *Leptopterigynandrum*, *Lescuraea*, *Leskeella*, *Lindbergia*, and *Rozea* were located in different main clades (at node 31, 28, 25, and 19).

*Incongruence and interaction among data sets*

*ILD test*

The *P*-values for all four data sets were analyzed by paired ILD test. The result was shown in Table 3. Significant discordances (*P* value < 0.05) were found in all combinations except rps4/atpB-rbcL (*P* = 0.053).

*PABA approach*

Thirty-three selected nodes were shown in Fig.2. The bootstrap support for each node was listed in Table 4 (all bootstrap support ≤5% were set to 5). The Mean δ values for each node at different data partitions were calculated following the method of Struck *et al*. (2006) and Struck (2007) and the result was included in Table 5. (For some nodes with no bootstrap alteration, the mean δ values were set to "N. A.". The averages over all nodes were showed at the last line of Table 6. Move to figure legends)

For nodes (node1, 3, 4, 7, 8, 13, 16, 21, 31, 33) that show no signal conflict, they represent monophyletic clades. The analysis using trnL-F data set showed negative effects on some nodes (added as $2^{nd}$: 11/33, as $3^{rd}$: 8/33, as $4^{th}$:16/33), with the average varied between 6 to N.A. Similarly, the study using rps4 or atpB-rbcL resulted in negative values at multiple nodes, with the average varied between 7 to N.A. More positive contributions to reconstruction were presented in ITS date set. ITS date set showed negative effect only at node 2 and 5, with the average changed from 54 to 47. Within four date sets respectively, as more data was included, the positive effects tend to decrease, indicating the increased effect of other partitions.

Discussion

*Influence of data conflict and utility of molecular markers*

Because the chloroplast genome is uniparentally inherited as a unit and not subject to recombination, multiple cpDNA sequences and restriction sites can be readily combined (Soltis & Soltis 2000). Consequently, chloroplast DNA *rbcL*, *rps4*, *trnL-F*, and *atpB-rbcL* became popular sequences for early phylogeny reconstruction of pleurocarpous mosses. However, we still found incongruence between three sequences used in our study. In paired ILD test of combinations (*trnL-F/rps4*, *trnL-F/atpB-rbcL*, and *rps4/atpB-rbcL*), only *P*-value (0.053) indicated congruence between *rps4/atpB-rbcL*, while incongruence presented in other two pairs. In PABA approach, three cpDNA sequences showed conflict nodes among themselves, e. g. node 5, 10, 19, and so on. On the other hand, even if *trnL-F*, *rps4*, and *atpB-rbcL* were used to analyses separately or simultaneously, the resolution was very insufficient on the backbone of Hypnales with very low bootstrap supports. For example the bootstrap at nodes (19/20, 25/26, 28/29, 31/32) all was 5% (see in table 4). The foundational root was the high level of homoplasy resulting from the rapid diversification, and chloroplast sequences had not enough date sites for reconstruction above the level of genus in Hypnales. Therefore, more sequences with rapid evolution were needed to be added into reconstruction.

Nuclear ITS was introduced since it can more easily be aligned across genera or even families of pleurocarps than that in congeneric species of many other mosses (Shaw *et al*. 2002; Vanderpoorten *et al*. 2002). It was not necessary to worry too many variations in sequences to analyse. Indeed, the average levels (i.e. $2^{nd}$ 54, $3^{rd}$ 48, $4^{th}$ 47) of ITS on all nodes improved evidently with respect to three chloroplast sequences since there were more parsimony-informative site (824) in ITS than that (433) in total three chloroplast sequences. In this study, support of the backbone mainly came from ITS sequence, while three cpDNA sequences only provided little support even negative effect. On the other hand, since its biparental inheritance, ITS sequence maybe complicated the stable relationships within genus supported by cpDNA sequences, e. g. the resolution about genus *Thuidium*.

*Phylogenetic relationships of families and taxonomic implications*
*Monophyly of Thuidiaceae*

In all topologies of all combinations and analyses, node 16 was supported stably. The bootstrap support of the node changed from 5 to 100 (showed in table 4) in all combinations of PABA, and four sequences all show nonnegative influence without conflict signal among them (in table 5). Species in node 16 formed a monophyletic group including *Abietinella*, *Actinothuidium*, *Bryonoguchia*, *Pelekium*, *Haplocladium*, *Helodium*, *Leskea*, *Platylomella*, *Rauiella*, and *Thuidium*.

*Haplocladium* was a genus of Anomodontoideae (Brotherus 1925) and then of Thuidieae (Watanabe 1972; Wu & Jia 2000; Wu *et al*. 2002) in traditional Thuidiaceae. In later three phylogenies (Buck & Goffinet 2000; Goffinet & Buck 2004; Goffinet & *et al*. 2008), it was transferred with *Claopodium* to Leskeaceae. In this study, three *Haplocladium* species, *Leskea scabrinervis*, and *Platylomella lescurii* form a clade (node 12) though conflict signals existed at this node. In Fig.4.b.1A of Troisky et al. (2007), some species of Leskeaceae aligning with *Thuidium*, *Rauiella*, *Abietinella*, and *Helodium* formed a clade (bs: 91), which included *Leskea polycarpa*, *L. gracukescens*, *Haplocladium virginianum*, *H. anustifolium* and *Pseudoleskeopsis zippelii*. Garc á-Avila *et al*. (2009) showed the similar result that *Leskea polycarpa* and *Haplocladium micorphyllum* aligning with *Rauiella* and *Abietinella* also formed a clade (bs: 60). According Vanderpoorten *et al*. (2003), *Platylomella lescurii* appeared nested within the Thuidiaceae/Leskeaceae, while it aligned assuredly within *Haplocladium* in our study.

Other species of Leskeaceae in our study including *Lindbergia sinensis*, *Lindbergia serrulatus*, *Regmatodon declinatus*, *Rozea chrysea*, *Leptopterigynandrum incurvatum*, *Leptopterigynandrum subintegrum*, *Leskeella nervosa*, *Lescuraea mutabilis* separately belonged to different main four or five clades in Fig. 2 & 3. There was no monophyletic Leskeaceae in all combinations. Other families such as Amblystegiaceae, Anomodontaceae, and Hypnaceae were also polyphyletic groups. These results conformed to general conclusions by Troitsky *et al*. (2007) and Olsson *et al*. (2009).

Three Helodiaceae genera (i.e. *Actinothuidium*, *Bryonoguchia*, and *Helodium*), *Abietinella*, and *Rauiella* formed a monophyletic clade (at node 14). Though main support came from ITS sequence, this result was similar to the definition for subfam. Helodioideae by Wu & Jia (2000) and Wu *et al*. (2002) except that gen. *Rauiella* had to be transferred from Thuidioideae to align with other four genera. Additionally, some evident characters in Thuidiaceae, e. g. pinnate branches, markedly paraphylliate axis, and papillose laminal cells, present simultaneously in these genera. It is more proper to resolve these genera under subfam. Helodioideae.

We support a Thuidiaceae *s. lat.* be recognized. This group shares a common ancestor, although it is unique only if the Helodiaceae and *Haplocladium* (Leskeaceae) are taken into it.

*The monophyly of Thuidium and Pelekium*

Concerning the division of *Thuidium* and *Pelekium*, there were two main viewpoints. Warnstorf

(1905), Watanabe (1972), Buck & Crum (1990), Wu & Jia (2000), Touw (2001) and Wu *et al.* (2002) recognized them as two different genera. In their opinion, it is difficult to make clear subgeneric division. Alternatively, Brotherus (1925) following Fleischer (1923) divided *Thuidium* into two subgenera (i.e. *Thuidiella* and *Euthuidium*). Touw (1976), Chen (1978), and Fang & Koponen (2001) also remained these two subgenera (i.e. *Thuidium* and *Microthuidium*). The species of subgenus *Thuidium* usually survive in thick mats, accompanied by a dioecous sexual system; paraphyllia diverse in form, at least 10 cells long, extensively forked. Comparatively, subgenus *Microthuidium* commonly occurs in thin mats, matched by an autoecous one; paraphyllia filiform, mostly fewer than 10 cells long, usually unforked. Though presenting different treatments, it was a reconcilable conflict according to our results. *Thuidium* and *Pelekium* were resolved as monophyletic group respectively. *Thuidium* clade and *Pelekium* clade further composed of a clade (i.e. node 12) with relatively high support (bs: 72, bpp: 1.00). This result is consistent with the two clades reported by Soares (2015). These two clades could be accepted as two genera or two subgenera. Given the significant differences between the two clades in terms of life form and sexual strategy, we suggest maintaining them as two separate genera.

*The character evolution in Thuidiaceae s. lat.*

The main gametophytic characteristics of Thuidiaceae are: (1) stem usually regularly or irregularly branched, pinnate to tetra-pinnate; (2) paraphyllia present or dense, forked or unforked; (3) papillae over laminal cells, unipapillae to pluripapillae; and (4) single strong costa in most species (all in our study). We divided our Thuidiaceae *s. lat.* into three groups (A, B, & C in Fig.2 & 3). From A to C, the types of branching, paraphyllia, and papillae changed from simple to complex. The plants of A are irregularly and pinnately branched or pinnately branched, and plants of B are regularly pinnately (for most species) to bi-pinnately branched, and plants of C are regularly bi- to tripinnately branched. The paraphyllia are present or rare on stems of part A in contrast with abundant or dense of parts B and C, and are also less variable than later two in paraphyllium morphology and leaf dimorphism. The stem leaves of A and B have unipapillose in the center or at the upper corner of each lumen, while part C have abundant phenotypic diversities especially in *Thuidium*, e.g. unipapillae, stellate papillae, scattered pluripapillae, horseshoe-shaped papillae, and mixed or irregular papillae. To study these varieties will lead to a better understanding of character evolution of the Thuidiaceae.

In further analysis of group C (*Thuidium* & *Pelekium*), four distinct groups (CI to CIV) were distinguished (Fig.2). In *Pelekium* (CI & CII) and *Thuidium* (CIII & CIV), similar evolution patterns of characters were presented on paraphyllium cells (especially on the apical cell) and laminal cells. The number of papilla decreased from CI to CII and from CIII to CIV. In *Pelekium* clade, the paraphyllium apical cells of CI (from *P. tamariscellum* to *P. gratum*) are usually truncate with 1-3 papillae, while that of CII (*P. haplohymenium*, *P. minusculum* & *P. fuscatum*) are usually acute without papilla. The median laminal cells of CI are centrally unipapillose, while that of CII are pluripapillose. On *Thuidium* laminal cells, the stellate papilla (in *T. pristocalyx*) and horseshoe-shaped papilla (in *T. subglaucinum*) resemble a pluripapillose condition (Fang & Koponen 2001). The median paraphyllium and laminal cells of CIII (from *T. pristocalyx* to *T. subglaucinum*) are pluripapillose, while that of CIV (from *T. cymbifolium* to *T. philibertii*) are unipapillose. The paraphyllium apical cells of CIV are unipapillose, while that of CIII are pluripapillose (usually ≥ 4 papillae). In terms of the evolution of microscopic characterizations ,

both of these genera have undergone a simplified process.

Conclusion

We dare not declare that our results about the relationships within Thuidiaceae are conclusive completely since more taxa are needed in the analyses. However, our methods give a chance to settle the high levels of homoplasy in Hypnales. More molecular evidences to corroborate the relationships within pleurocarpous mosses may be selected in nuclear sequences like ITS and ETS since, in general, the chloroplast and mitochondrial genome evolves more slowly than nuclear genome in plant (Wolfe *et al*. 1987). On the other hand, cpDNA sequences in this study were relatively suited to reconstruct relationships within genus of pleurocarpous mosses. Further studies about derivation and evolution of Thuidiaceae, even of pleurocarpous mosses can be advanced by the presence of well-supported phylogenetic hypotheses.

Acknowledgments

We thank the assistance of many collaborators. We thank Jia Heng and Zhou Nan-Nan for help in collecting specimens; we thank Prof. Wu Peng-Chen (Institute of Botany, The Chinese Academy of Sciences), Prof. Cao Tong (ShangHai Normal University) and Prof. Wang You-Fang (East China Normal University) for the identification of specimens. We also wish to thank Dr. Gong Lei in our laboratory for molecular biological experiments. We are grateful to Torsten H. Struck (Department of Biology Chemistry, University of Osnabrüeck, Germany) for proposals about the analyses and providing software for PABA analyses. The project was supported by the Doctorate Fellowship Foundation of Nanjing Forestry University and National Natural Sciences Foundation of China (No. 41761103, 30270110 & 30070155).

References

Brotherus, V. F. (1925) Musci (Laubmoose) II. In: Engler, A. & Prantl, K. (eds.), Die Natürlichen Pflanzenfamilien. W. Engelmann, Leipzig. pp. 1–522.

Buck, W. R., Goffinet, B. & Show. A. J. (2000) Novel relationships in pleurocarpous mosses as revealed by cpDNA sequences. The Bryologist 103(4): 774–789.

Buck, W. R. & Goffinet, B. (2000) Morphology and classification of mosses (Bryopsida). In: Shaw, A.J. & Goffinet, B. (eds.), Bryophyte Biology. Cambridge Univ. Press, Cambridge, pp. 71–123.

Buck, W. R. & Crum, H. A. (1990) An evaluation of familial limits among the genera traditionally aligned with the Thuidiaceae and Leskeaceae. Contr. Univ. Mich. Herb. 17: 55–69.

Chen, P.-C. (1978) Genera Muscorum Sinicorum . Sci. Press, Beijing. pp. 1–331.

Chiang, T.-Y., Barbara, A. S. & Peng, C.-I. (1998) Universal primers for amplification and sequencing a noncoding spacer between the atpB and rbcL genes of chloroplast DNA. Bot. Bull. Acad. Sin. 39: 245–250.

Crum, H. A. & Anderson, L. E. (1981) Mosses of eastern North America. Vol. 2. Columbia Univ. Press, New York. pp. 665–1328.

García-Avila, D., De Luna, E. & Newton A. E. (2009) Phylogenetic relationships of the Thuidiaceae and the non-monophyly of the Thuidiaceae and the Leskeaceae based on rbcL, rps4 and the rps4-trnS intergenic spacer. The Bryologist 112(1): 80–93.

Gardiner, A., Ignatov, M. S., Huttunen, S. & Troitsky, A.V. (2005) Pseudoleskeaceae and Pylaisiaceae. On resurrection of the families Pseudoleskeaceae. Schimp. and Pylaisiaceae. Schimp. (Musci, Hypnales). Taxon 54(3): 651–663.

Goffinet, B. & Buck, W. R. (2004) Systematics of Bryophyta: from molecules to a revised classification. Monograp hs in Systematic Botany from the Missouri Botanical Garden, 98: 205–239.

Goffinet, B., Buck, W. R. & Shaw, A. J. (2008) Morphology and classification of the Bryophyta. Pages 55–138. In: B. Goffinet & A.J. Shaw (eds.), Bryophyte Biology, 2$^{nd}$ edition. Cambridge University Press.

Guindon, S. & Gascuel O. (2003) A simple, fast, and accurate algorithm to estimate large phylogenies by maximum likelihood. Syst. Biol. 52: 696–704.

Fang, Y.-M. & Koponen, T. (2001) A revision of *Thuidium*, *Haplocladium*, and *Claopodium* (Musci, Thuidiaceae) in China. Bryobrothera 6: 1–81.

Farris, J. S., Kallersjo, M., Kluge, A. G. & Bult, C. (1995) Constructing a significance test for incongruence. Syst. Biol. 44: 570 –572.

Fleischer, M. (1904–1923) Die Musci der Flora von Buitenzorg (Zugleich Laubmoosflora von Java), Vols. 1–4. Brill, Leiden.

Hall, T. (2005) BioEdit version 7.0.5. Available from http://www.mbio.ncsu.edu/BioEdit/bioedit.html.[Online.]

Hedenäs, L. (1994) The basal pleurocarpous diplolepidous mosses – a cladistic approach. The Bryologist 97: 225–243.

Hedenäs, L. (1995) Higher taxonomic level relationships among diplolepidous pleurocarpous mosses – a cladistic overview. Journal of Bryology 18: 723–781.

Hedenäs, L. (1997) An evaluation of phylogenetic relationships among the Thuidiaceae, the Amblystegiaceae, and the temperate members of the Hypnaceae. Lindbergia 22: 101–133.

Hedenäs, L. (1998) Cladistic studies on pleurocarpous mosses: Research needs, and use of results. In Bates J. W., N. W. Ashton, and J. G. Duckett (Eds.), Bryology for the Twenty-first century, W. S. Maney & Son and the Br. Bryol. Soc., Leeds. pp. 125–141.

Huelsenbeck, J. P. & Ronquist , F. (2001) MRBAYES: Bayesian inference of phylogeny. Bioinformatics 17: 754–755.

Huttunen, S., Bell, N., Bobrova, V.K., Buchbender, V., Buck, W.R., Cox, C.J., Goffinet, B., Hedenäs, L., Ho, B.-C., Ignatov, M.S., Krug, M., Kuznetsova, O., Milyutina, I.A., Newton, A., Olsson, S., Pokorny, L., Shaw, J.A., Stech, M., Troitsky, A., Vanderpoorten, A., Quandt, D. (2012) Disentangling knots of rapid evolution: origin and diversification of the moss order Hypnales. J. Bryol. 34, 187–211.

Huttunen, S., Ignatov, M.S., Quandt, D., Hedenäs, L. (2013) Phylogenetic position and delimitation of the moss family Plagiotheciaceae in the order Hypnales. Bot. J. Lin. Soc. 171, 330–353.

Kindberg, N. C. (1897) Leskeaceae. In Genera of European and North American Bryineae (mosses): Synoptically Disposed. Linkoeping, Sweden. pp. 9–59.

Kropp, B. R., Hansen, D. R., Wolf, P. G., Flint, K. M. & Thomson, S. V. (1997) A Study on the Phylogeny of the Dyer's Woad Rust Fungus and Other Species of *Puccinia* from Crucifers. Phytopathology 87(5): 565–571.

Newton, A.E., Wikström, N., Bell, N., Forrest, L.L. & Ignatov, M.S. (2007) Dating the diversification of the pleurocarpous mosses. In: A.E. Newton & R. Tangney, eds. *Pleurocarpous mosses: systematics and evolution. Systematics Association Special Volume Series* 71: 337–66.

Nadot, S., Bajon, R. & Lejeune, B. (1994) The chloroplast gene rps4 as a tool for the study of Poaceae phylogeny. Plant Systematics and Evolution 191: 27–38.

Olsson, S., V. Buchbender, J. Enroth, L. Hedenäs, S. Huttunene & Quandt, D. (2009) Phylogenetic analyses reveal high levels of polyphyly among pleurocarpous lineages as well as novel clades. The Bryologist 112(3): 447–466.

Page, R.D.M. (1996) Treeview: an application to display phylogenetic trees on personal computers. Comput. Appl. Biosci. 12: 357–358.

Posada, D. (2008) jModelTest: Phylogenetic Model Averaging. Molecular Biology and Evolution 25: 1253-1256.

Ronquist, F. & Huelsenbeck, J. P. (2003) MRBAYES 3: Bayesian phylogenetic inference under mixed models. Bioinformatics 19: 1572–1574.

Ronquist, F., Huelsenbeck, J. P. & Mark, P. van der. (2007) MrBayes 3.1 Manual from http://mrbayes.scs.fsu.edu/wiki/index.php/Manual. [Online.]

Schimper, W. P. (1876) *Synopsis Muscrum europaeorum*, ed. 2. Sumptibus Librariae E. Schweizerbart, Stdutgartiae.

Shaw, A. J., McDaniel, S. F., Warner, O. & Ros, R. M. (2002) New frontiers in bryology and lichenology. Phylogeography and phylodemography. The Bryologist 105: 373–383.

Shaw, A. J., Cox, C. J., Goffinet, B., Buck, W. R. & Boles, S. B. (2003) Phylogenetic evidence of a rapid radiation of pleurocarpous mosses (Bryophyta). Evolution 57(10): 2226–2241.

Soares, A.E.R. (2015) A Família Thuidiaceae Schimp. no Brasil, um estudo taxonômico, filogenético e morfológico. PhD Thesis, Universidade de Brasília.

Soltis, E. D. & Soltis,P. S. (2000) Contributions of plant molecular systematics to studies of molecular evolution. Plant Molecular Biology 42: 45–75.

Souza-Chies, T., Bittar, T. G., Nadot, S., Carter, L., Besin, E. & Lejeune, B. (1997) Phylogenetic analysis of *Iridaceae* with parsimony and distance methods using the plastid gene *rps*4. Plant Systematics and Evolution 204: 109–123.

Stech, M. & Frahm, J.-P. (1999) The status of Platyhypnidium mutatum Ochyra & Vanderpoorten and the systematic value of the Donrichardsiaceae based on molecular data. Journal of Bryology 21: 191–195.

Stech, M. & Quandt, D. (2010) 20,000 species and five key markers: The status of molecular bryophyte phylogenetics. Phytotaxa 9: 196–228.

Struck, T. H., Purschke, G., Halanych, K. M. (2006) Phylogeny of Eunicida (Annelida) and exploring data congruence using apartition addition bootstrap alteration (PABA) approach. Syst. Biol. 55: 1–20.

Struck, T. H. (2007) Data congruence, paedomorphosis and salamanders. Frontiers in Zoology 4: 22.

Swofford, D. L. (2002) PAUP*. Phylogenetic Analysis Using Parsimony (* and other methods). Version 4 (beta 10). Sinauer Associates, Sunderland, Massachusetts.

Taberlet, P., Gielly, L., Pautou G. & Bouvet ,J. (1991) Universal primers for amplification of three non-coding regions of chloroplast DNA. Plant Molecular Biology 17: 1105–1109.

Tamura, K., Dudley, J., Nei M. & Kumar, S. (2007) MEGA4: Molecular Evolutionary Genetics

Analysis (MEGA) software version 4.0. Molecular Biology and Evolution 24: 1596–1599.

Souza-Chies, T.T., Bittar, G., Nadot, S., Carter, L., Besin, E. &Lejeune, B. (1997) Phylogenetic analysis of Iridaceae with parsimony and distance methods using the plastid gene rps4. Plant Systematics and Evolution 204: 109–123.

Thompson, J. D., Gibson, T. J., Plewniak, F., Jeanmougin, F. & Higgins, D.G. (1997) The ClustalX windows interface: flexible strategies for multiple sequence alignment aided by quality analysis tools. Nucleic Acids Research 24: 4876–4882.

Touw, A. (1976) A taxonomic revision of *Thuidium*, *Pelekium*, and *Rauiella* (Musci: Thuidiaceae) in Africa South of the Sahara. Lindbergia 3: 135–195.

Touw, A. (2001) A review of the Thuidiaceae (Musci) and realignment of taxa traditionally accommodated in Thuidium sensu amplo (Thuidium Schimp., Thuidiopsis (Broth.) M. Fleisch., and Pelekium Mitt.), including Aequatoriella gen. nov. and Indothuidium gen. nov. J. Hattori Bot. Lab. 90: 167–209.

Troitsky, A. V., Ignatov, M. S., Bobrova, V. K. & Milyutina. I. A. (2007) Contribution of genosystematics to current concepts of phylogeny and classification of bryophytes. Biochemistry (Moscow) 72(12): 1368–1376.

Vanderpoorten, A., Hedenäs, L., Cox, C. J. & Shaw, A.J. (2002) Circumscription, classification, and taxonomy of Amblystegiaceae (Bryopsida) inferred from nuclear and chloroplast DNA sequence data and morphology. Taxon 51(1): 115–122.

Vanderpoorten, A., Goffinet, B., Hedenäs, L., Cox, C. J. & Shaw, A. J. (2003) A taxonomic reassessment of the Vittiaceae (Hypnales, Bryopsida): evidence from phylogenetic analyses of combined chloroplast and nuclear sequence data. Plant Systematics and Evolution 241: 1–12.

Warnstorf, C. (1905) Kryptogamenflora der Mark Brandenburg. Laubmoose. 2. Bontraeger, Berlin.

Watanabe, R. (1972) A revision of the family Thuidiaceae in Japan and adjacent areas. J. Hattori Bot. Lab. 36: 171–320.

Wolfe, K.H., Li, W.-H. & Sharp, P. M. (1987) Rates of nucleotide substitution vary greatly among plant mitochondrial, chloroplast, and nuclear DNAs. Proc Natl Acad Sci USA 84:9054–9058.

Wu, P.-C., Wang, M.-Z. & Zhong, B.-G. (2002)Thuidiaceae. In Wu, P.-C. (eds.), Moss flora of China 6. Hookeriaceae–Thuidiaceae. Sci. Press, Beijing. pp. 191–270.

Wu, P.-C. & Jia, Y. (2000) A revision of Chinese Thuidiaceae (s. l., Musci). Acta Phytotaxonomica Sinica 38: 256–265.

Table 1. The taxonomy of broad sense Thuidiaceae

| Brotherus (1925) | Goffinet et al. (2008) | Watanabe (1972) For Japan & adjacent area | Wu (2002) For China |
|---|---|---|---|
| **Anomodontoideae** | **Anomodontaceae** | **Anomodontoideae** | **Anomodontaceae** |
| *Anomodon* | *Anomodon* | *Anomodon* | *Anomodon* |
| *Claopodium* | *Bryonorrisia* | *Haplohymenium* | *Haplohymenium* |
| *Haplocadium* | *Chileobryon* | *Herpetineuron* | *Herpetineuron* |
| *Haplohymenium* | *Haplohymenium* | *Miyabea* | *Miyabea* |
| *Herpetineuron* | *Herpetineuron* | | **Thuidiaceae** |
| *Miyabea* | *Miyabea* | **Heterocladioideae** | **Heterocladioideae** |
| **Helodioideae** | *Schwetschkeopsis* | *Heterocladium* | *Heterocladium* |
| *Actinothuidium* | | *Leptopterigynandrum* | *Leptocladium* |
| *Bryonoguchia* | **Helodiaceae** | | *Leptopterigynandrum* |
| *Helodium* | *Actinothuidium* | | **Helodioideae** |
| *Hylocomiopsis* | *Bryochenea* | **Thuidioideae** | *Abietinella* |
| **Heterocladioideae** | *Helodium* | *Abietinella* | *Actinothuidium* |
| *Heterocladium* | | *Actinothuidium* | *Bryonoguchia* |
| *Leptopterigynandrum* | **Thuidiaceae** | *Boulaya* | *Helodium* |
| **Thuidioideae** | *Abietinella* | *Bryonoguchia* | **Thuidioideae** |
| *Abietinella* | *Boulaya* | *Claopodium* | *Boulaya* |
| *Boulaya* | *Cyrto-hypnum* | *Haplocadium* | *Claopodium* |
| *Orthothuidium* | *Fauriella* | *Helodium* | *Cyrto-hypnum* |
| *Pelekium* | *Pelekium* | *Hylocomiopsis* | *Haplocadium* |
| *Rauiella* | *Rauiella* | *Pelekium* | *Pelekium* |
| *Thuidiopsis* | *Thuidiopsis* | *Rauiella* | *Rauiella* |
| *Thuidium* | *Thuidium* | *Thuidium* | *Thuidium* |

Table 2. Taxa and GenBank Accession numbers in this study

| Taxa | *trnL-F* | *rps4-trnS* | *atpB-rbcL* | **ITS** |
|---|---|---|---|---|
| *Abietinella abietina* | AY009850 | AY907953 | AF322308 | AY009802 |
| *Actinothuidium hookeri* | KF770502 | KF770556 | KF770610 | KF770664 |
| *Amblystegium serpens* | FJ535739 | FJ572627 | FJ535758 | FJ535778 |
| *Anomodon giraldii* | KF770518 | KF770572 | KF770626 | KF770680 |
| *Anomodon minor* | KF770515 | KF770569 | KF770623 | KF770677 |
| *Antitrichia curtipendula* | AY010286 | AY908570 | EU186641 | DQ974370 |
| *Brachythecium salebrosum* | FJ572448 | FJ572653 | AY663309 | GQ246855 |
| *Brotherella erythrocaulis* | KF770528 | KF770582 | KF770636 | KF770690 |
| *Bryonoguchia molkenboeri* | KF770504 | KF770558 | KF770612 | KF770666 |
| *Calliergon cordifolium* | AY857569 | AF469844 | AY857590 | AY857611 |
| *Calliergonella cuspidate* | GQ428068 | AY908183 | AF322310 | AF168145 |
| *Claopodium assurgens* | KF770524 | KF770578 | KF770632 | KF770686 |
| *Claopodium pellucinerve* | KF770523 | KF770577 | KF770631 | KF770685 |
| *Cyrto-hypnum bonianum* | KF770492 | KF770546 | KF770600 | KF770654 |
| *Cyrto-hypnum contortulum* | KF770496 | KF770550 | KF770604 | KF770658 |
| *Cyrto-hypnum fuscatum* | KF770488 | KF770542 | KF770596 | KF770650 |
| *Cyrto-hypnum gratum* | KF770491 | KF770545 | KF770599 | KF770653 |
| *Cyrto-hypnum haplohymenium* | KF770490 | KF770544 | KF770598 | KF770652 |
| *Cyrto-hypnum minusculeum* | KF770489 | KF770543 | KF770597 | KF770651 |
| *Cyrto-hypnum sparsifolium* | KF770495 | KF770549 | KF770603 | KF770657 |
| *Cyrto-hypnum tamariscellum* | KF770494 | KF770548 | KF770602 | KF770656 |
| *Cyrto-hypnum vestitissimum* | KF770493 | KF770547 | KF770601 | KF770655 |
| *Entodon cladorrhizans* | KF770510 | KF770564 | KF770618 | KF770672 |
| *Haplocladium angustifolium* | KF770497 | KF770551 | KF770605 | KF770659 |
| *Haplocladium microphyllum* | KF770498 | KF770552 | KF770606 | KF770660 |
| *Haplocladium sp.* | KF770500 | KF770554 | KF770608 | KF770662 |
| *Haplocladium strictulum* | KF770499 | KF770553 | KF770607 | KF770661 |
| *Haplocladium virginianum* | AF161133 | AF143040 | AF322305 | AF168160 |
| *Haplohymenium triste* | KF770514 | KF770568 | KF770622 | KF770676 |
| *Helodium blandowii* | AY009852 | AY908339 | AF322313 | AY009803 |
| *Helodium paludosum* | KF770505 | KF770559 | KF770613 | KF770667 |
| *Herpetineuron toccoae* | KF770519 | KF770573 | KF770627 | KF770681 |
| *Homalothecium sericeum* | AF397805 | DQ294319 | EF531013 | EF617596 |
| *Homomallium connexum* | KF770511 | KF770565 | KF770619 | KF770673 |
| *Hookeria acutifolia* | AY306763 | AF306929 | AF413569 | FM161137 |
| *Hygrohypnum smithii* | AY857565 | AY908620 | AY857586 | AY857607 |
| *Hylocomium splendens* | KF770513 | KF770567 | KF770621 | KF770675 |
| *Hypnum hamulosum* | KF770527 | KF770581 | KF770635 | KF770689 |
| *Juratzkaea sinensis* | KF770522 | KF770576 | KF770630 | KF770684 |
| *Leptopterigynandrum austro-alpinum* | KF770526 | KF770580 | KF770634 | KF770688 |
| *Leptopterigynandrum tenellum* | KF770525 | KF770579 | KF770633 | KF770687 |
| *Lescuraea mutabilis* | KF770520 | KF770574 | KF770628 | KF770682 |
| *Lescuraea mutabilis(1)* | AY683601 | AY663326 | AY663291 | AY737456 |
| *Leskea scabrinervis* | KF770501 | KF770555 | KF770609 | KF770663 |
| *Leskeella nervosa* | KF770517 | KF770571 | KF770625 | KF770679 |
| *Leucodon sciuroides* | KF770529 | KF770583 | KF770637 | KF770691 |
| *Lindbergia serrulatus* | KF770508 | KF770562 | KF770616 | KF770670 |
| *Lindbergia sinensis* | KF770506 | KF770560 | KF770614 | KF770668 |
| *Myuroclada maximoviczii* | KF770521 | KF770575 | KF770629 | KF770683 |
| *Neckera pennata* | AF315072 | AF143008 | AF322357 | AY009809 |
| *Okamuraea hakoniensis* | KF770516 | KF770570 | KF770624 | KF770678 |
| *Platydictya jungermannioides* | AY857568 | AF469833 | AY857589 | AY857610 |
| *Platyhypnidium riparioides* | DQ208209 | AY908298 | AY857595 | FJ476003 |
| *Platylomella lescurii* | AY683601 | AY663326 | AY663291 | AY737456 |
| *Pylaisiella selwynii* | KF770512 | KF770566 | KF770620 | KF770674 |
| *Rauiella fujisana* | KF770503 | KF770557 | KF770611 | KF770665 |
| *Regmatodon declinatus* | KF770507 | KF770561 | KF770615 | KF770669 |
| *Rhytidium rugosum* | FJ572454 | FJ572612 | EU186639 | FJ572383 |
| *Rozea chrysea* | KF770509 | KF770563 | KF770617 | KF770671 |
| *Sanionia uncinata* | FJ572455 | FJ572608 | AF322321 | AF168148 |
| *Scorpidium scorpioides* | AY626014 | AY908584 | AY625977 | AY625995 |
| *Thuidium assimile* | KF770481 | KF770535 | KF770589 | KF770643 |
| *Thuidium cymbifolium* | KF770482 | KF770536 | KF770590 | KF770644 |
| *Thuidium delicatulum* | KF770478 | KF770532 | KF770586 | KF770640 |
| *Thuidium glaucinoides* | KF770487 | KF770541 | KF770595 | KF770649 |
| *Thuidium kanedae* | KF770484 | KF770538 | KF770592 | KF770646 |
| *Thuidium philibertii* | KF770476 | KF770530 | KF770584 | KF770638 |
| *Thuidium philibertii* | KF770479 | KF770533 | KF770587 | KF770641 |
| *Thuidium pristocalyx* | KF770486 | KF770540 | KF770594 | KF770648 |
| *Thuidium sp.* | KF770477 | KF770531 | KF770585 | KF770639 |
| *Thuidium subglaucinum* | KF770480 | KF770534 | KF770588 | KF770642 |
| *Thuidium subglaucinum* | KF770483 | KF770537 | KF770591 | KF770645 |

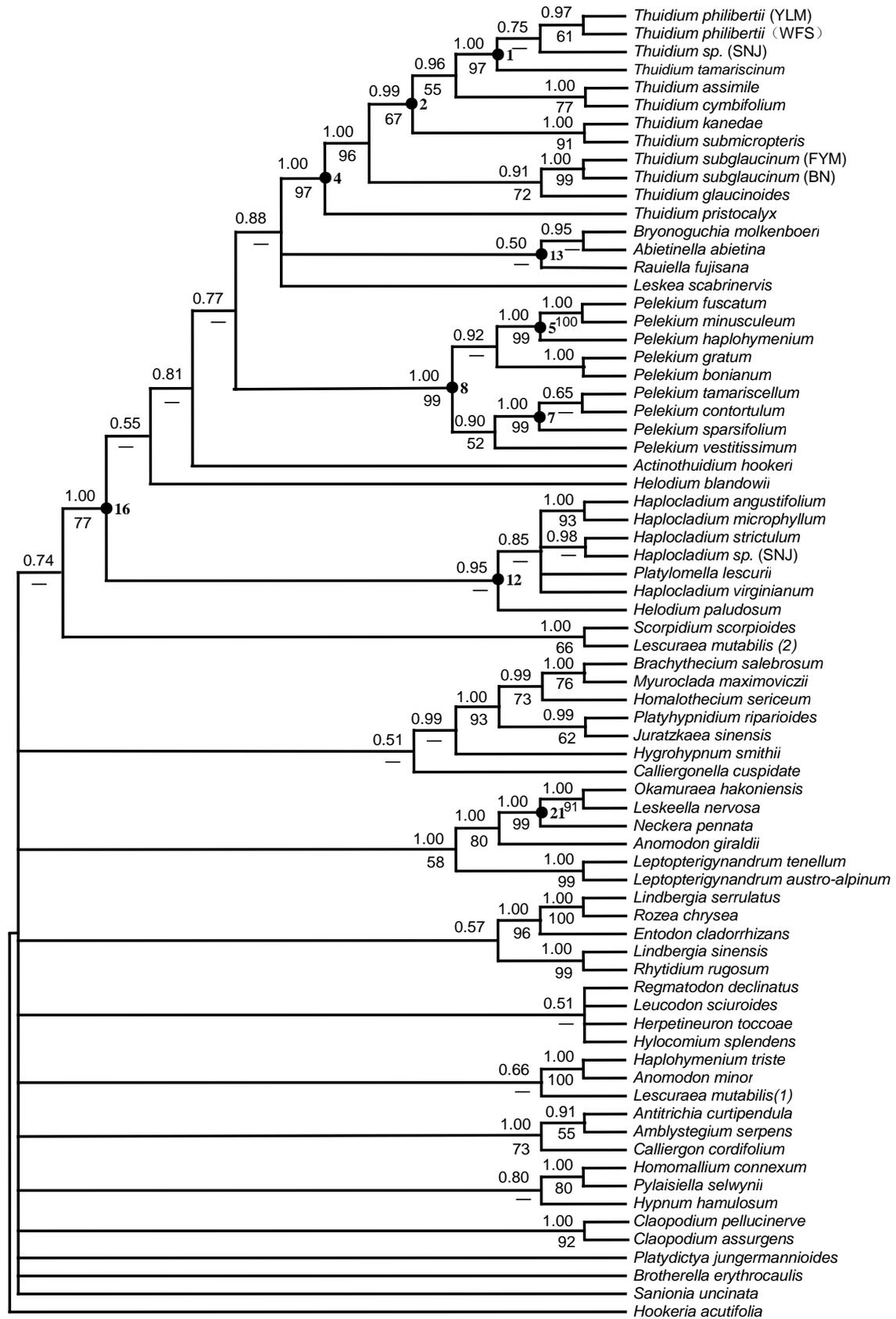

Fig.1. Major-rule consensus BI tree based on trnL-F, rps4, & atpB-rbcL. Bayesian posterior probabilities (bpp), (>0.50) is indicated above the branches; bootstrap support (bs) (>50%) of MP analysis under the branches.

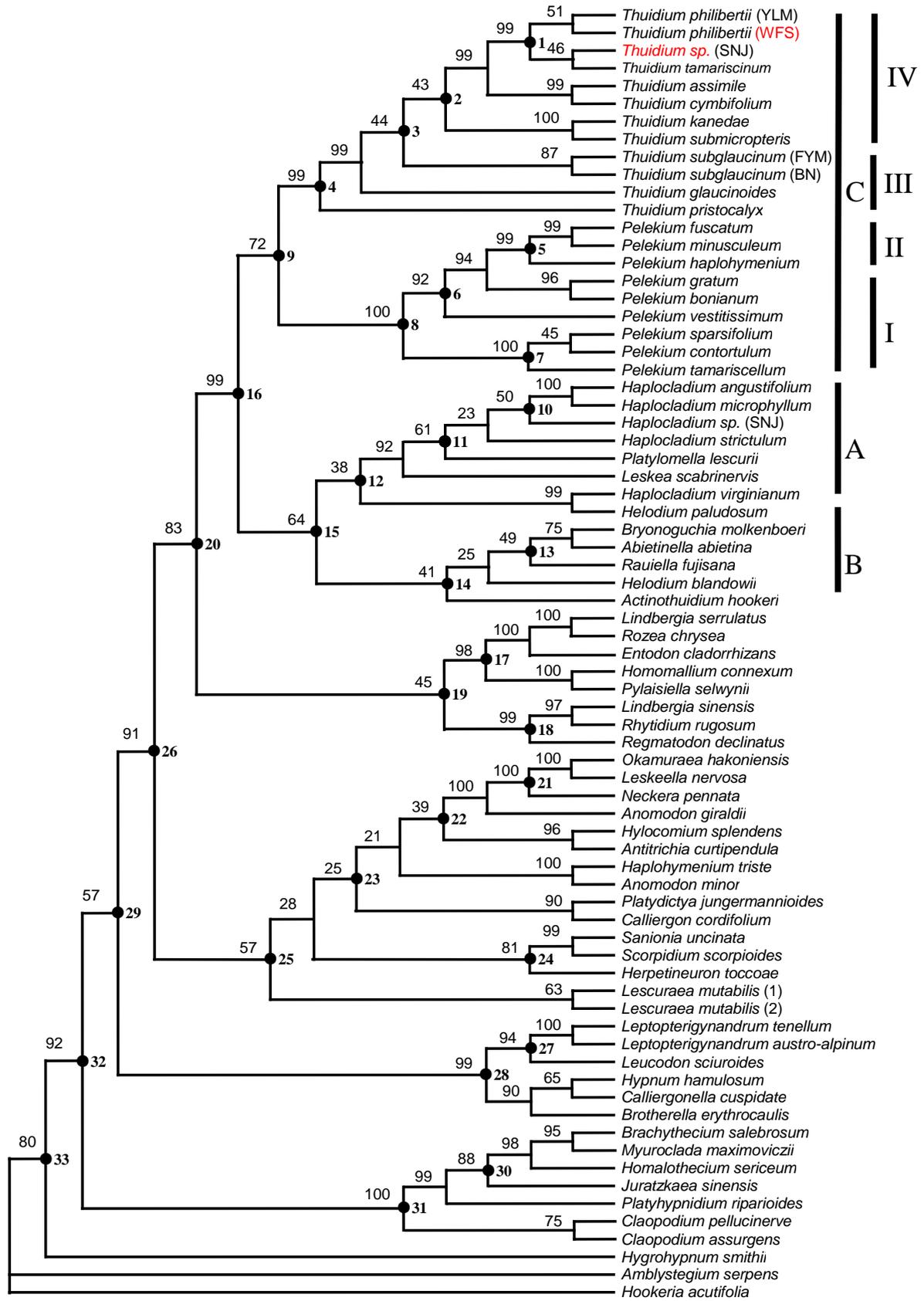

Fig.2. MP tree based on combination of trnL-F, rps4, atpB-rbcL, & ITS. Bootstrap support (bs) is indicated above the branches.

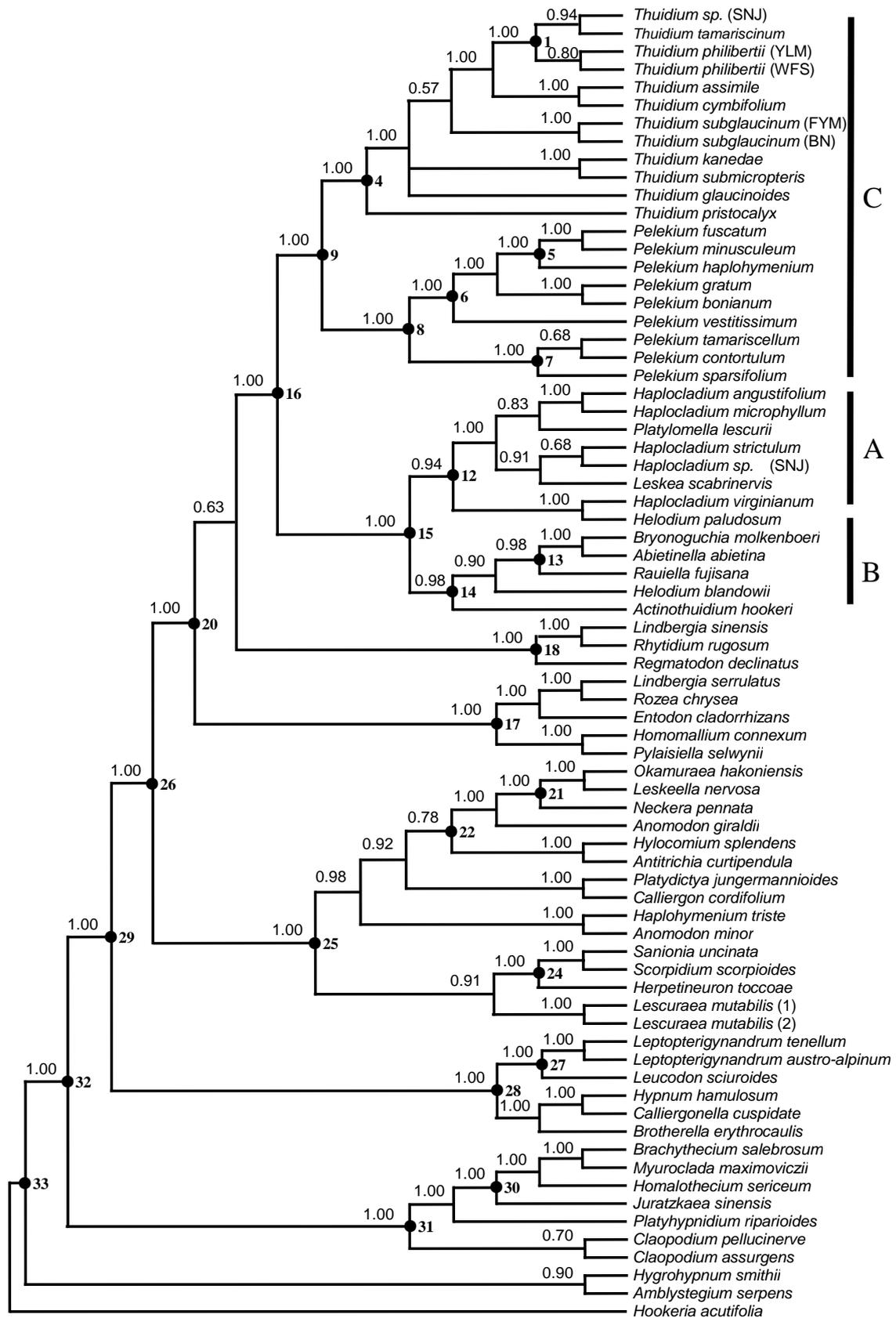

Fig. 3 Major-rule consensus BI tree based on combination of trnL-F, rps4, atpB-rbcL, & ITS. Bayesian posterior probabilities (bpp) (>50%) is indicated above the branches.

Table 3 Results of the ILD test of congruence of data sets

| against | rps4-trnS | atpB-rbcL | ITS |
|---|---|---|---|
| trnL-F | 0.024* | 0.001* | 0.001* |
| rps4-trnS | – | 0.053 | 0.004* |
| atpB-rbcL | | – | 0.001* |

(*Significant discordances between partitions, $P$ values < 0.05)

Table 4 Bootstrap values in all four date sets corresponding to the nodes (bipartitions) of combined analyses of all genes.

| Node | Date set | | | | | | | | | | | | | |
|---|---|---|---|---|---|---|---|---|---|---|---|---|---|---|
| | ABCD | A | B | C | D | AB | AC | AD | BC | BD | CD | ABC | ABD | ACD | BCD |
| 1 | 99 | 63 | 70 | 79 | 57 | 89 | 92 | 87 | 92 | 86 | 77 | 98 | 96 | 96 | 95 |
| 2 | 43 | 5 | 5 | 34 | 14 | 47 | 34 | 18 | 70 | 30 | 30 | 67 | 30 | 33 | 42 |
| 3 | 44 | 0 | 10 | 6 | 5 | 9 | 5 | 23 | 6 | 13 | 10 | 7 | 36 | 33 | 25 |
| 4 | 99 | 82 | 88 | 27 | 73 | 99 | 85 | 93 | 71 | 89 | 97 | 98 | 98 | 100 | 99 |
| 5 | 99 | 5 | 100 | 90 | 96 | 98 | 80 | 94 | 100 | 100 | 99 | 100 | 100 | 99 | 100 |
| 6 | 92 | 10 | 5 | 10 | 100 | 5 | 59 | 99 | 5 | 95 | 99 | 15 | 95 | 99 | 93 |
| 7 | 100 | 86 | 98 | 48 | 100 | 100 | 93 | 100 | 99 | 100 | 100 | 100 | 100 | 100 | 100 |
| 8 | 100 | 60 | 39 | 88 | 100 | 77 | 99 | 97 | 97 | 98 | 100 | 99 | 98 | 100 | 100 |
| 9 | 72 | 5 | 5 | 5 | 85 | 33 | 5 | 87 | 5 | 86 | 73 | 10 | 85 | 73 | 76 |
| 10 | 50 | 5 | 5 | 31 | 67 | 5 | 8 | 39 | 25 | 69 | 73 | 8 | 43 | 50 | 78 |
| 11 | 61 | 5 | 5 | 6 | 60 | 5 | 10 | 44 | 6 | 63 | 72 | 9 | 48 | 58 | 77 |
| 12 | 38 | 5 | 5 | 5 | 32 | 5 | 5 | 28 | 5 | 31 | 46 | 5 | 25 | 40 | 41 |
| 13 | 49 | 5 | 5 | 5 | 5 | 5 | 15 | 18 | 6 | 5 | 17 | 18 | 19 | 49 | 16 |
| 14 | 41 | 5 | 5 | 5 | 46 | 5 | 5 | 44 | 5 | 47 | 50 | 5 | 43 | 42 | 51 |
| 15 | 64 | 5 | 5 | 5 | 80 | 5 | 5 | 78 | 5 | 80 | 70 | 5 | 74 | 68 | 68 |
| 16 | 99 | 37 | 5 | 15 | 90 | 60 | 62 | 99 | 43 | 95 | 96 | 78 | 100 | 100 | 98 |
| 17 | 98 | 5 | 5 | 5 | 99 | 11 | 5 | 100 | 5 | 99 | 98 | 5 | 99 | 98 | 97 |
| 18 | 99 | 5 | 13 | 5 | 100 | 5 | 5 | 99 | 11 | 100 | 100 | 11 | 100 | 100 | 100 |
| 19 | 45 | 5 | 5 | 5 | 18 | 5 | 5 | 21 | 5 | 15 | 41 | 5 | 22 | 49 | 35 |
| 20 | 83 | 5 | 5 | 5 | 61 | 5 | 5 | 68 | 5 | 75 | 76 | 5 | 77 | 75 | 87 |
| 21 | 100 | 66 | 95 | 41 | 100 | 97 | 91 | 100 | 96 | 100 | 100 | 100 | 100 | 100 | 100 |
| 22 | 39 | 5 | 5 | 5 | 49 | 5 | 5 | 42 | 5 | 51 | 47 | 5 | 44 | 41 | 48 |
| 23 | 25 | 5 | 5 | 5 | 35 | 5 | 5 | 27 | 5 | 34 | 34 | 5 | 12 | 26 | 32 |
| 24 | 81 | 5 | 5 | 5 | 89 | 5 | 5 | 93 | 5 | 90 | 77 | 5 | 92 | 84 | 76 |
| 25 | 57 | 5 | 5 | 5 | 70 | 5 | 5 | 65 | 5 | 69 | 75 | 5 | 61 | 60 | 74 |
| 26 | 91 | 5 | 5 | 5 | 90 | 5 | 5 | 93 | 5 | 90 | 90 | 5 | 92 | 90 | 90 |
| 27 | 94 | 5 | 5 | 6 | 91 | 5 | 5 | 92 | 5 | 87 | 95 | 5 | 90 | 96 | 93 |
| 28 | 99 | 5 | 5 | 5 | 100 | 5 | 5 | 100 | 5 | 100 | 100 | 5 | 100 | 100 | 100 |
| 29 | 57 | 5 | 5 | 5 | 59 | 5 | 5 | 56 | 5 | 53 | 61 | 5 | 54 | 61 | 60 |
| 30 | 88 | 23 | 25 | 7 | 46 | 63 | 27 | 84 | 8 | 82 | 68 | 35 | 97 | 87 | 71 |
| 31 | 100 | 5 | 5 | 5 | 100 | 5 | 5 | 100 | 5 | 100 | 100 | 5 | 100 | 100 | 100 |

| 32 | 92 | 5 | 5 | 5 | 92 | 5 | 5 | 94 | 5 | 91 | 92 | 5 | 91 | 93 | 93 |
| 33 | 80 | 5 | 5 | 5 | 35 | 5 | 5 | 42 | 5 | 41 | 74 | 5 | 47 | 78 | 77 |

Table 5 Alteration of bootstrap supported to important nodes in Fig. 2 as data partitions are added. (N.A. = not applicable due to no alteration)

| Node | BS value | Gene | | | | | | | | | | | | |
|---|---|---|---|---|---|---|---|---|---|---|---|---|---|---|
| | | trnL-F as | | | rps4-trnS as | | | atpB-rbcL as | | | ITS as | | | |
| | | 2nd | 3rd | 4th | 2nd | 3rd | 4th | 2nd | 3rd | 4th | 2nd | 3rd | 4th | |
| 1 | 99 | 21 | 12 | 4 | 23 | 11 | 3 | 24 | 9 | 3 | 13 | 5 | 1 | |
| 2 | 43 | 15 | N.A. | N.A. | 31 | 19 | 10 | 37 | 16 | 13 | 11 | -15 | -24 | |
| 3 | 44 | 5 | 16 | 19 | 4 | 10 | 11 | N.A | 7 | 8 | 8 | 25 | 37 | |
| 4 | 99 | 30 | 13 | N.A. | 26 | 7 | -1 | 3 | 5 | 1 | 27 | 14 | 1 | |
| 5 | 99 | -5. | N.A. | -1 | 36 | 9 | N.A. | 26 | 2 | -1 | 33 | 7 | -1 | |
| 6 | 92 | 16 | 3 | -1 | -5 | -18 | -7 | 16 | 3 | -3 | 89 | 73 | 77 | |
| 7 | 100 | 16 | N.A. | N.A. | 22 | 2 | N.A. | 3 | N.A. | N.A. | 23 | 3 | N.A. | |
| 8 | 100 | 15 | 1 | N.A. | 8 | N.A. | N.A. | 32 | 9 | 2 | 36 | 8 | 1 | |
| 9 | 72 | 10 | 1 | -4 | 10 | 2 | -1 | -4 | -16 | -13 | 77 | 64 | 62 | |
| 10 | 50 | -17 | -22 | -28 | -1 | 3 | 0 | 10 | 8 | 7 | 47 | 44 | 42 | |
| 11 | 61 | -4 | -9 | -16 | 1 | 3 | 3 | 6 | 11 | 13 | 55 | 54 | 52 | |
| 12 | 38 | -1 | -4 | -3 | N.A. | -3 | -2 | 5 | 7 | 13 | 30 | 30 | 33 | |
| 13 | 49 | 8 | 19 | 33 | N.A. | 2 | 0 | 8 | 18 | 30 | 8 | 19 | 31 | |
| 14 | 41 | -1 | -4 | -10 | N.A. | N.A. | -1 | 1 | 1 | -2 | 42 | 40 | 36 | |
| 15 | 64 | -1 | -3 | -4 | N.A. | -2 | -4 | -3 | -7 | -10 | 71 | 65 | 59 | |
| 16 | 99 | 37 | 15 | 1 | 19 | 7 | N.A. | 23 | 7 | N.A. | 78 | 44 | 21 | |
| 17 | 98 | 2 | N.A. | 1 | 2 | -1 | N.A. | N.A. | -3 | -1 | 94 | 91 | 93 | |
| 18 | 99 | -3 | N.A. | -1 | 2 | 2 | -1 | -1 | 2 | -1 | 92 | 93 | 88 | |
| 19 | 45 | 1 | 5 | 10 | -1 | -2 | -4 | 8 | 16 | 23 | 21 | 35 | 40 | |
| 20 | 83 | 2 | N.A. | -4 | 5 | 7 | 5 | 5 | 6 | 5 | 68 | 75 | 78 | |
| 21 | 100 | 17 | 1 | N.A. | 29 | 3 | N.A. | 9 | 1 | N.A. | 33 | 5 | N.A. | |
| 22 | 39 | -2 | -4 | -9 | 1 | 1 | -2 | -1 | -1 | -5 | 42 | 39 | 34 | |
| 23 | 25 | -3 | -11 | -7 | N.A. | -6 | -1 | N.A. | -1 | 13 | 27 | 18 | 20 | |
| 24 | 81 | 1 | 3 | 5 | N.A. | -1 | -3 | -4 | -8 | -11 | 82 | 79 | 76 | |
| 25 | 57 | -2 | -8 | -17 | N.A. | -2 | -3 | 5 | -3 | -4 | 65 | 60 | 52 | |
| 26 | 91 | 1 | 1 | 1 | N.A. | N.A. | 1 | N.A. | -1 | -1 | 86 | 86 | 86 | |
| 27 | 94 | N.A. | 1 | 1 | -2 | -1 | -2 | 1 | 3 | 4 | 86 | 88 | 89 | |
| 28 | 99 | N.A. | N.A. | -1 | N.A. | N.A. | -1 | N.A. | N.A. | -1 | 95 | 95 | 94 | |
| 29 | 57 | -1 | N.A. | -3 | -2 | -1 | -4 | 1 | 1 | 3 | 52 | 53 | 52 | |
| 30 | 88 | 32 | 20 | 17 | 26 | 8 | 1 | 3 | -12 | -9 | 60 | 52 | 53 | |
| 31 | 100 | N.A. | N.A. | N.A. | N.A. | N.A. | N.A. | N.A. | N.A. | N.A. | 95 | 95 | 95 | |
| 32 | 92 | 1 | N.A. | -1 | N.A. | -1 | -1 | N.A. | N.A. | 1 | 87 | 87 | 87 | |
| 33 | 80 | 2 | 3 | 3 | 2 | 3 | 2 | 13 | 24 | 33 | 47 | 62 | 75 | |
| Average over all nodes | | 6 | 1 | N.A. | 7 | 2 | N.A. | 7 | 3 | 3 | 54 | 48 | 47 | |